\documentclass[aps,pra,amsmath,twocolumn,amssymb,footinbib,superscriptaddress,showpacs]{revtex4-1}

\usepackage{amssymb,amsfonts,amsmath}

\usepackage{color}

\usepackage[apple mac]{inputenc}
\usepackage[english]{babel}
\usepackage{times}
\usepackage{latexsym}
\usepackage{fancyhdr}
\usepackage{verbatim}
\usepackage{graphicx}

\usepackage{epsf}
\usepackage{subfigure}
\usepackage{bm}
\usepackage{epstopdf}

\definecolor{bblue}{rgb}{0.,0.24,0.51}
\newcommand{\blue}{\color{blue}}
\newcommand{\ket}[1]{\ensuremath{| #1 \rangle}}
\newcommand{\bra}[1]{\ensuremath{\langle #1 |}}

\begin{document}

\title{{\blue Realistic Quantum Simulation of the Topological Mott Insulator}}
\title{Quantum Simulation of a Topological Mott Insulator with Rydberg Atoms in a Lieb Lattice}

\author{A. Dauphin}
\email{adauphin@ulb.ac.be}
\affiliation{Center for Nonlinear Phenomena and Complex Systems - Universit\'e Libre de Bruxelles , 231, Campus Plaine, B-1050 Brussels, Belgium}\affiliation{Departamento de F\'isica Te\'orica I, Universidad Complutense, 28040 Madrid, Spain}

\author{ M. M\"uller}
\affiliation{Departamento de F\'isica Te\'orica I, Universidad Complutense, 28040 Madrid, Spain}
\author{M. A. Martin-Delgado}
\affiliation{Departamento de F\'isica Te\'orica I, Universidad Complutense, 28040 Madrid, Spain}

\begin{abstract}
We propose a realistic scheme to quantum simulate the so-far experimentally unobserved topological Mott insulator phase -- an interaction-driven topological insulator -- using cold atoms in an optical Lieb lattice. To this end, we study a system of spinless fermions in a Lieb lattice, exhibiting repulsive nearest and next-to-nearest neighbor interactions, and derive the associated zero temperature phase diagram within mean-field approximation. In particular, we analyze how the interactions can dynamically generate a charge density wave ordered, a nematic as well as a topologically non-trivial quantum anomalous Hall phase. We characterize the topology of the different phases by the Chern number and discuss the possibility of phase coexistence. Based on the identified phases, we propose a realistic implementation of this model using cold Rydberg-dressed atoms in an optical lattice. The scheme, which allows one to access in particular the topological Mott insulator phase, robustly and independently of its exact position in parameter space, merely requires global, always-on off-resonant laser coupling to Rydberg states and is feasible with state-of-the-art experimental techniques that have already been demonstrated in the laboratory. 
\end{abstract}

\date{\today}

\pacs{37.10.Jk, 32.80.Rm, 71.10.Fd, 03.65.Vf, 73.43.Nq}

\maketitle

Topological phases of matter are attracting increasing interest not only from a fundamental standpoint, where they represent a new paradigm of quantum states which evade a classification by the standard Landau theory of phases \cite{wen-book,bernevig_2013}, but also due to their potential applications in quantum information processing \cite{Kitaev_2002,nayak_2008}. Here, in particular topological insulators (TIs), being insulating in the bulk, but exhibiting robust, topologically protected current-carrying edge states, are a new subclass of quantum materials with topological properties \cite{Hasan_2010,Qi_2011,Moore_2010}. Their existence has been confirmed in pioneering experiments \cite{Konig_2007,Hsieh_2008,Hsieh_2009} after their theoretical prediction \cite{Kane_2005,Bernevig_2006,Fu_2007,Fu_2007b} in two and three dimensions.  

The discovery and study of topological insulators raise fundamental questions, in particular calling for an understanding of the various mechanisms that can lead to topological order and the classification of topological phases according to the symmetries of the system. For example, the presence of background gauge fields is one setting in which topological order can arise, and an extensive classification in the absence of interactions has been achieved \cite{Haldane_1988,kitaev2009,Ryu_2010,Teo_2010}. Alternatively, such phases can be generated in periodically driven quantum systems, giving rise to Floquet topological insulators \cite{Kitagawa_2010,Rudner_2013}. In this work, we focus on yet another type, the topological Mott insulator \cite{Hohenadler_2013}, in which the topologically non-trivial phase is generated dynamically by fermionic interactions, even in the complete absence of background gauge fields. Such phase has been first predicted theoretically for fermions in a honeycomb lattice geometry with nearest and next-to-nearest neighbor interactions in a seminal work by Raghu \textit{et al.} \cite{Raghu_2008}. Subsequently, this mechanism has been explored in other lattice geometries, including so far checkerboard \cite{Sun_2009}, Kagome \cite{Sun_2009,Wen_2010,Liu_2010}, Lieb \cite{Tsai_2015}, modified dice \cite{Dora_2014} and decorated honeycomb \cite{Wen_2010} lattices, as well as by refined mean-field theory \cite{Weeks_2010,Grushin_2013} and exact diagonalization studies \cite{Garcia_Martinez_2013,Daghofer_2014,djuri__2014}. These showed that the location and width of the parameter region where this phase is predicted to occur crucially depends on the lattice geometry, filling, interaction pattern and size of the system. Whereas most of these works, and in particular a recent numerical study \cite{djuri__2014}, provide growing evidence for the existence of such a topological Mott insulating phase, the experimental observation of this predicted novel quantum phase is outstanding to date. 

\begin{figure}[t]
\begin{center}
	\includegraphics[scale=1]{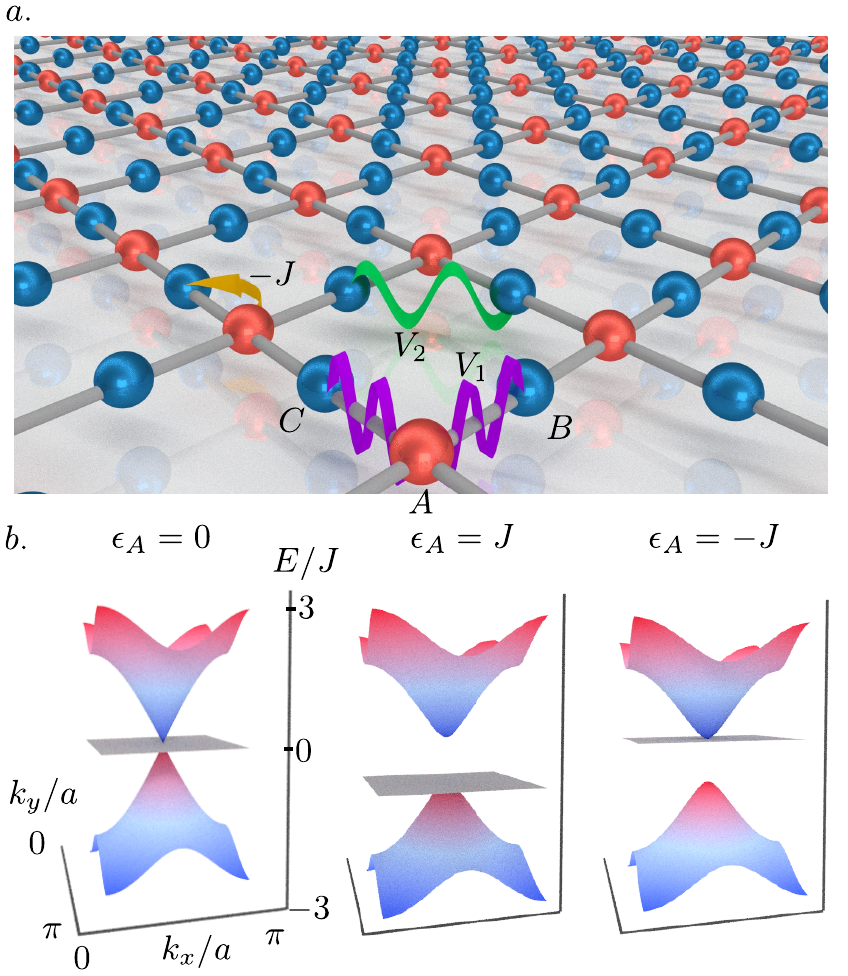}
	\caption{a. The Lieb lattice is a square lattice of $N_\Lambda$ unit cells with a three-atom unit cell A,B,C. The Hamiltonian has hopping between nearest neighbors and interactions between nearest neighbors and next-to-nearest neighbors. b. Energy spectrum of the non-interacting Lieb lattice with $\epsilon_A=0$ (left subfigure). The system exhibits one Dirac cone at $k_x=k_y=\pi/2$. The two right subfigures present the energy spectrum for $\epsilon_A=-J$ (respectively $\epsilon_A=J$): the flat band is still present but the spectrum has now an energy gap between band one and two (respectively two and three).}
\label{fig:fig1}
\end{center}
\end{figure}

Here, we propose and work out a realistic scheme which provides a route to an experimental observation of a topological Mott insulator phase using cold interacting fermionic Rydberg atoms in an optical lattice. An earlier implementation proposal \cite{Dauphin_2012} has been followed by complementary works which have considered the creation of a TMI phase in the continuum \cite{Kitamura_2015} as well as the occurrence of topological density waves in cold atomic gases \cite{Li_2015}. The present work is motivated by recent enormous progress that has been made in using cold atoms in optical lattices as a platform to explore and study in a controlled way topological quantum phases of matter. In particular, topological phases in square and honeycomb lattices or band structures with topological features have been demonstrated recently, based on the engineering of background gauge fields, both for bosonic  \cite{Aidelsburger_2013,Miyake_2013,Aidelsburger_2014} as well as more recently also fermionic atoms \cite{Jotzu_2014}. Complementary,  highly flexible, tunable optical lattice setups allowing one to implement e.g.~hexagonal lattice geometries as well as to imprint external staggering potentials have been demonstrated in various laboratories  \cite{Tarruell_2012,Uehlinger_2013,Jotzu_2014}. In particular, recently an optical Lieb lattice \cite{Taie_2015} (see Fig.~\ref{fig:fig1}a) as well as Lieb photonic lattices \cite{Vicencio_2015,Guzman_2014,Mukherjee_2015} have been realized experimentally. In addition, the ingredients for the realization of extended Hubbard models in optical lattices, in particular tunable, long-range interactions by off-resonant laser coupling of ground state atoms to highlying Rydberg states \cite{Nath-2010, Pupillo-2010,Balewski-2014} have been demonstrated in a series of recent experiments \cite{Viteau-2011,Schauss-2012,Schauss-2015,Weber-2015,Labuhn_2015}. 

In this work, on the one hand we quantitatively study a model of interacting fermions on a Lieb lattice, which hosts a topologically non-trivial Mott insulator phase in a broad parameter region of the phase diagram; on the other hand we provide realistic implementation schemes which are feasible with the above-mentioned state-of-the-art experimental techniques that have been already demonstrated individually, and partially also in a combined way, in the laboratory. Among the distinguishing features of our work are: (i) The interaction pattern is implementation-friendly, in particular, the topological phase exists in the naturally accessible regime of moderately large long-range interactions: it involves strong next-to-nearest neighbor interactions $V_2$, which however are still allowed to be smaller than nearest-neighbor interactions $V_1$ ($V_2 \lesssim V_1)$. Note that the requirement of dominant next-to-nearest neighbor interactions $V_2  > V_1$ constitutes a considerable obstacle for the implementation of many models \cite{Raghu_2008,Weeks_2010,Wen_2010,Dauphin_2012}. Furthermore, the required interactions between fermions at different lattice sites correspond to a global, translationally invariant pattern. (ii) The proposed scheme requires control over effectively spinless fermions moving in a 2D Lieb lattice, for which previous works have shown how to realize such an optical lattice geometry by an appropriate combination of standing wave laser fields \cite{Shen_2010,Apaja_2010}. (iii) We remark that in the implementation scheme we propose all Hamiltonian couplings are independently controllable, and no fine tuning is required, neither of experimental implementation parameters nor in view of the derived (mean-field) phase diagram, in which the non-trivial phase exists in a wide parameter stripe. Specifically, we will present two possible variants for an implementation: In the first, fermionic atoms in the Lieb lattice are exposed to an external staggering potential, and only globally applied, spatially homogeneous Rydberg laser dressing is required. In the second scheme, no staggering potential is needed, interactions $V_1$ and $V_2$ are tunable completely independently, and the Rydberg laser dressing profile is applied according to a spatially periodic pattern, which can be realized by a superimposed 2D square optical lattice, imposing moderately higher control requirements. 

We consider a system of spinless fermions on the Lieb lattice with repulsive nearest and next-to-nearest neighbor interactions (see Fig.~\ref{fig:fig1}a), described by the second-quantized Hamiltonian
\begin{equation}
\label{eq:hamint}
H=H_0+V_1\sum_{\langle i,j \rangle}n_i n_j+V_2\sum_{\langle \langle i,j \rangle \rangle}n_i n_j\text{,}
\end{equation}
where $n_i=c^\dagger_i c_i$ is the onsite number operator. The tight-binding Hamiltonian
\begin{equation}
H_0=-J\sum_{\langle i,j \rangle}c^\dagger_j c_i +\epsilon_A\sum_i c^\dagger_ic_i \delta_{A,i}
\end{equation}
describes hopping between nearest-neighbor sites with a real-valued amplitude $-J$, and it involves an onsite external staggering potential $\epsilon_A$, which accounts for a difference in chemical potential between $A$ and $B,C$ sites. The energy spectrum can be obtained analytically and is composed of three energy bands: a flat band at $E=0$ and two dispersive bands $E_\pm(\mathbf{k})=\epsilon_A/2\pm \sqrt{(\epsilon_A/2)^2+4J^2(\cos^2 k_x+\cos^2 k_y)}$ \cite{Weeks_2010b}. For vanishing staggering potential $\epsilon_A=0$, the system has one Dirac cone at $k_x=k_y=\pi/2$ with a linear dispersion relation in its vicinity (see Fig.~\ref{fig:fig1}b). At the Fermi energy $E_F=0$, the system is semi-metallic. For $\epsilon_A \neq 0$, the spectrum still possesses a flat band but an energy gap appears in the energy spectrum and the dispersion relation around the Dirac point becomes quadratic (see Fig.~\ref{fig:fig1}b). The position of the energy gap depends of the sign of $\epsilon_A$. For the Fermi energy lying in the energy gap, the system is in a trivial insulating phase. The latter can be verified by computing the transverse conductivity $\sigma_{xy}=\sum_i C_i \sigma_0$, which is proportional to the sum of the Chern numbers $C_i$ of the occupied bands and to the conductivity quantum $\sigma_0$ \cite{Thouless_1982}, yielding $\sigma_{xy}=0$.

We treat the interacting part of the Hamiltonian within a mean-field approximation, according to the Hartree-Fock decoupling \cite{Bruus_2004} of the quartic interaction term, $n_in_j \simeq - \xi_{ij} c^\dagger_j c_i - \xi_{ij}^* c^\dagger_i c_j +\vert \xi_{ij} \vert^2 
+\bar{n}_i c_j^\dagger c_j+ \bar{n}_j c^\dagger_i c_i - \bar{n}_i \bar{n}_j$. This decoupling gives rise to 

nearest neighbor hopping terms proportional to the expected value $\xi_{ij}=\langle c^\dagger_i c_j \rangle$ and to onsite terms proportional to the expected value $\bar{n}_i=\langle n_i \rangle$. We then determine the mean field phase diagram by solving self-consistently the set of coupled equations for the expected values $\xi_{ij}$ and $\bar{n}_i$. To this end, two natural assumptions are made to limit the number of order parameters: translation invariance of the system, allowing one to work with the three-atom basis, and isotropy of the hopping between $A$ and $B,C$ and between $B$ and $C$ sites, enabling one to work only with $\xi_{AB}=\xi_{AC}$ and $\xi_{BC}$.

To characterize the physics of the different phases, we introduce several order parameters, written in terms of $\bar{n}_i$ or $\xi_{ij}$. The charge density wave (CDW) order parameter $\rho_1=\bar{n}_A-\bar{n}_B-\bar{n}_C$ characterizes the difference of occupation between sites $A$ on the one hand and $B,C$ sites on the other hand. The nematic order parameter $\rho_2=\bar{n}_B-\bar{n}_C$ accounts for a possible difference of occupation between $B$ and $C$ sites. In addition to these order parameters, the mean-field decoupling of the nearest neighbor interaction gives rise to a  renormalisation of the nearest neighbor hopping amplitude $J'=J+\xi_{AB}V_1$. Similarly, the next-to-nearest neighbor interactions generate a real-valued next-to-nearest neighbor hopping $\xi_{BC}^\text{R}$ as well as an imaginary-valued next-to-nearest neighbor hopping $\xi^\text{I}_{BC}$. The latter is responsible for the occurrence of a quantum anomalous Hall (QAH) phase \cite{Haldane_1988,Weeks_2010b} and constitutes a local order parameter for this phase (see App.~\ref{app:mfh} for details). Solving the self-consistent equations shows that the parameters $\xi_{AB}$ and $\xi_{BC}^\text{R}$ are almost constant as functions of $V_1$ and $V_2$, which justifies fixing their values to $\xi_{AB}=0.24J$ and $\xi_{BC}^\text{R}=-0.1J$ (see App.~\ref{app:nnre} for details).

The resulting mean field Hamiltonian reads then
\begin{equation}
\label{eq:hamintmf}
H_\text{MF}=E_0+\sum_{\mathbf{k}}\Psi^\dagger_\mathbf{k} \mathcal{H}_\mathbf{k} \Psi_\mathbf{k}\text{,}
\end{equation}
where $E_0/N_\lambda=4V_2 (\xi_{BC}^\text{I})^2-V_1/2(n^2-\rho_1^2)-V_2(1/4 n^2-\rho_2^2)$ is a real-valued energy shift depending only on the expectation values of the operators, $\Psi_\mathbf{k}^\dagger=(c^\dagger_{A,\mathbf{k}},c^\dagger_{B,\mathbf{k}},c^\dagger_{C,\mathbf{k}})$ and the matrix elements of $\mathcal{H}_\mathbf{k}$ are given by
\begin{equation}
\begin{split}
&\mathcal{H}_{\mathbf{k}}^{AA}=(n-\rho_1)V_1\text{, } \\
&\mathcal{H}_{\mathbf{k}}^{BB}=(n+\rho_1)V_1+(n-2\rho_2)V_2\text{, }  \\
&\mathcal{H}_{\mathbf{k}}^{CC}=(n+\rho_1)V_1+(n+2\rho_2)V_2\text{, } \\
&\mathcal{H}_{\mathbf{k}}^{AB}=-2J' \cos{k_x} \text{, } \qquad \mathcal{H}_{\mathbf{k}}^{AC}=-2J' \cos{k_y},\\
&\mathcal{H}_{\mathbf{k}}^{BC}=-4V_2 \xi_{BC}^\text{R} \cos{k_x}\cos{k_y}+4 V_2\xi_{BC}^\text{I}\text{i} \sin{k_x} \sin{k_y} \text{.}
\end{split}
\end{equation}
To obtain the mean field phase diagram, we minimize the free energy $F$  at zero temperature, 
\begin{equation}
F/N_\Lambda=\langle H_{MF} \rangle/N_\Lambda=E_0+\sum_{i,\mathbf{k}\in \text{B.Z.}} E_i(\mathbf{k})\Theta(E_F-E_i)\text{.}
\end{equation}

\begin{figure}[t]
\begin{center}
	\includegraphics[scale=1]{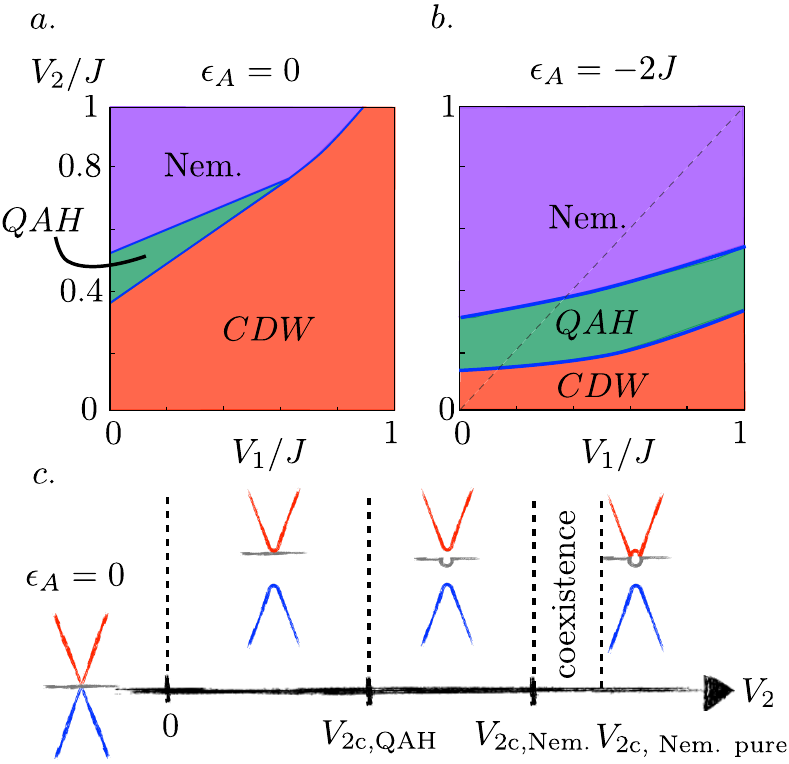}
	\caption{a. Phase diagram at a filling $2/3$ with $\epsilon_A=0$. Three phases are present: charge density wave (CDW), nematic (Nem.) and quantum anomalous Hall phase (QAH). The blue lines show the second order phase transitions. b. Phase diagram at $\epsilon_A=-2J$. The QAH is now stabilized in a bigger window and the boundaries have moved. c. Sketch of the energy spectrum of the line $V_1=0$ for $\epsilon_A=0$: When $V_2<V_{2\text{c, QAH}}=0.37J$, the lowest gap is opened by the $V_2$ interaction. For $V_2>V_{2\text{c, QAH}}$, the second energy gap opens and the system is in a QAH phase. For $V_2>V_{2\text{c, Nem.}}=0.47J$, the system is in the nematic phase; the second gap closes but the second energy band is still dispersive. For $V_{2\text{c, Nem. pure}}=0.52J$, the system is in a pure nematic phase.}
\label{fig:fig2}
\end{center}
\end{figure} 

We focus on the situation where the two first energy bands are filled (filling $n=2/3$). Note that in this case, even in the absence of a staggering potential ($\epsilon_A=0$), there develops a density imbalance between $A$ and $B,C$ sites, which becomes manifest through $\rho_1=-J$. 

We now discuss the mean field phase digram for $\epsilon_A=0$ (shown in Fig.~\ref{fig:fig2}a) and for clarity of the discussion first focus on the effect of the nearest neighbor interactions separately (horizontal axis for $V_2=0$). As soon as $V_1>0$, $\rho_1$ is renormalized and the system passes from a semimetallic (SM) phase to the CDW phase with an energy gap opening between the first and second band, while the gap between the second and third band still remains closed. 

Considering next exclusively the effect of the next-to-nearest neighbor interactions (vertical axis in Fig.~\ref{fig:fig2} for $V_1 =0$), a sketch of the different phases as a function of $V_2$ is shown in Fig.~\ref{fig:fig2}c. First, at small values of $V_2$, the next-to-nearest neighbor interaction gives rise to a CDW term in the mean field Hamiltonian proportional to $V_2 n$. The latter drives the system from the SM to the CDW phase by continuously opening an energy gap between the first and second energy band. Beyond a critical value, $V_2>V_{2\text{c, QAH}}$, $\xi^\text{I}_{BC}$ becomes nonzero and the system undergoes a second-order phase transition from the CDW to the QAH phase: the flat band becomes dispersive and an energy gap opens between the second and third band. We characterize the topological nature of the energy gap by computing the transverse conductivity, which is proportional to the sum of the Chern numbers of the occupied bands. We first compute numerically the Chern numbers of the two first energy bands with the gauge-invariant algorithm proposed by Fukui, Hastugai and Suzuki (FHS algorithm) \cite{Fukui_2005}. We find $C_1=0$, $C_2=1$ and hence $\sigma_{xy}=(C_1+C_2)\sigma_0=-1\times \sigma_0$, indicating the topological character of the phase. For $V_2>V_{2\text{c, Nem.}}$, $\rho_2$ starts to become nonzero, allowing the coexistence of the topological and nematic phases. There is a small parameter window of coexistence in which $\rho_2$ increases continuously while $\xi^\text{I}_{BC}$ decreases continuously. For $V>V_{2\text{c, Nem. pure}}$, $\xi^\text{I}_{BC}=0$, the second gap closes and the system enters a purely ungapped nematic phase characterized by a dispersive second energy band (see App.~\ref{app:detailsmf} for the plot of the order parameters in terms of $V_2$). Finally, we consider both nonzero $V_1$ and $V_2$ interactions and find a finite window (green region in Fig~\ref{fig:fig2}) where the system resides in the QAH phase.

Fig.~\ref{fig:fig2}b.~shows the phase diagram for the case of a finite external staggering potential, $\epsilon_A=-2J$. As shown in Fig.~\ref{fig:fig1}b., a negative value of $\epsilon_A$ opens the first energy gap of the non-interacting system. This term changes radically the phase diagram of the system with interactions. Focussing first exclusively on the next-to-nearest neighbor interaction (vertical axis), we find that the second order phase transition to the QAH phase appears for $V_{2c\text{, QAH}}=0.13J$, i.e.~at a much smaller critical value than for the case in Fig.~\ref{fig:fig2}a, however, with the size of the parameter region of the QAH phase of the same order. The transition to the nematic phase follows the same mechanism: the order parameter $\rho_2>0$ starts to grow for $V_2>V_{2c\text{, Nem.}}=0.27J$ and then $\xi^\text{I}_{BC}$ continuously decreases to zero giving rise to a purely ungapped nematic phase for $V_2>V_{2c\text{, Nem. pure}}=0.3J$. However, when considering the whole phase diagram with both nonzero $V_1$ and $V_2$ interactions, as a striking difference the region where the system is in the QAH phase is now a stripe: the fact that for each value of $V_1$, there is a value of $V_2$ such that the system is predicted to be in the QAH regime, will be important in the scheme we propose for quantum simulating this model.

For completeness, we briefly comment on the case of filling $1/3$. Here, for a vanishing staggering potential $\epsilon_A=0$, neither the QAH phase nor the nematic phase appear and the only phase generated by the interactions is the charge density wave ordered phase, corresponding to an opening of the energy gap between bands one and two. However, for finite $\epsilon_A>0$, the system can exhibit both QAH and nematic phase, and here $\epsilon_A$ counteracts the formation of the CDW phase.

\begin{figure}[h]
\begin{center}
	\includegraphics[scale=1]{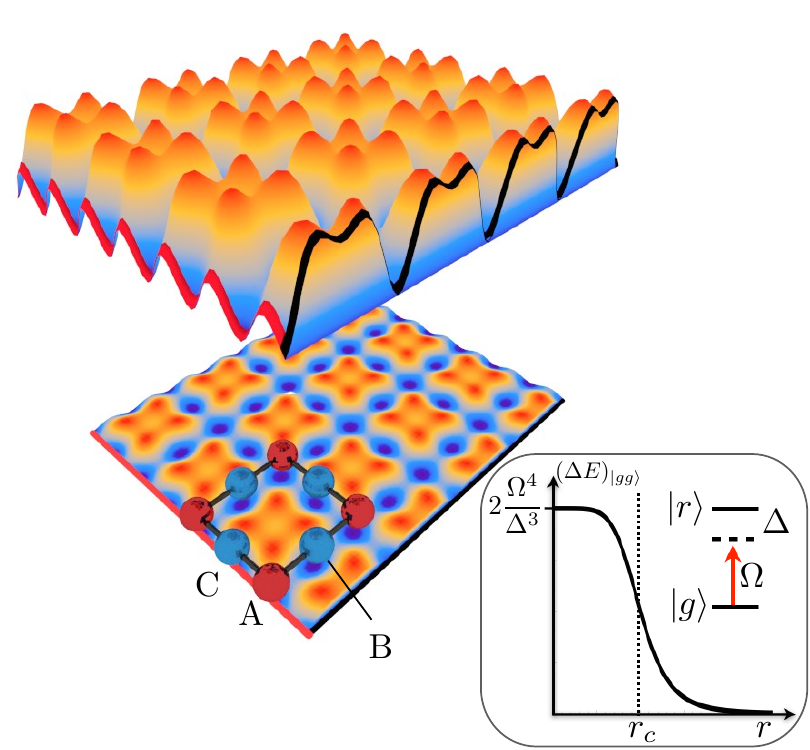}
	\caption{Upper subfigure: Optical lattice potential (a.u.) generated by three super-imposed lattice potentials, realizing the desired Lieb lattice structure ($U_0 = U_1 =2 U_2$) Red and black wavy lines show cuts of the potential landscape along the $x$-direction (crossing C and A sites) and $y$-direction (going through C-type lattice sites only). Lower part: projection of the potential landscape showing allowed lattice sites (blues wells) and energetically forbidden regions (yellow-reddish regions). Lower subfigure: Off-resonant laser coupling (Rabi frequency $\Omega$, detuning $\Delta$) leads to a weak admixture of Rydberg states $\ket{r}$ to the state of atoms residing in an electronic ground state $\ket{g}$, which induces an effective pairwise interaction potential as sketched in the right part.}
\label{fig:fig3}
\end{center}
\end{figure} 

For the implementation of the system as described by the Hamiltonian (\ref{eq:hamint}), we focus on cold fermionic and effectively spinless atoms loaded into an optical lattice exhibiting the Lieb lattice geometry as sketched in Fig.~\ref{fig:fig3}. This 2D lattice geometry in the xy-plane can be achieved, as shown in a recent experiment \cite{Taie_2015}, by superimposing three optical lattice potentials \cite{Shen_2010,Apaja_2010,Goldman_2011},
\begin{eqnarray}
U_{\text{OL}} & = & U_0 \left[ \sin^2 (kx) + \sin^2 (ky) \right] \\
& + & U_1 \left[ \sin^2 (2kx) + \sin^2 (2ky) \right] \nonumber \\
& + & U_2 \left[ \cos^2 \left(\sqrt{2}k \frac{x+y}{\sqrt{2}} \right) + \cos^2 \left( \sqrt{2}k \frac{x-y}{\sqrt{2}} \right) \right]. \nonumber
\end{eqnarray}
Tunnelling amplitudes as well as a staggering potential can be controlled by adjusting the relative intensities of the constituent lattice potentials: in particular, the choice $U_0 = U_1$ allows one to induce and control the staggering potential between A and B/C sites with a magnitude $\epsilon_A = U_0 - 2 U_2$ (see Fig.~\ref{fig:fig3} for the resulting potential and a possible choice of parameters). 

In order to induce the required repulsive interactions $V_1$ and $V_2$ (see Fig.~\ref{fig:fig3}) we propose to off-resonantly laser couple the fermionic electronic ground state atoms to highly excited Rydberg states, which exhibit strong and long-range repulsive interactions \cite{saffman-rmp-82-2313}. Such off-resonant laser dressing with Rydberg states, which has been recently demonstrated in the laboratory \cite{Schauss-2015, Weber-2015}, allows one to generate for experimentally realistic parameters effective repulsive nearest and nearest-neighbor interactions, which are on the same order of magnitude as the tunneling time scale $J$. In addition, this can be achieved in a regime where imperfections such as decay due to the finite lifetime of Rydberg states set in at sufficiently longer time scales (see App.~\ref{app:details_rydberg} for a more detailed analysis and a discussion of relevant error sources). In the regime of weak, off-resonant and red-detuned laser dressing, the effectively induced interaction potential exhibits the well-known plateau-like potential structure \cite{Nath-2010,Pupillo-2010} (see Fig.~\ref{fig:fig3}), with a nearly distance-independent interaction energy shift at small interparticle distances and rapidly decaying interactions beyond a critical interparticle distance $r_c$, which can be controlled by an appropriate choice of laser parameters and atomic Rydberg states (see App.~\ref{app:details_rydberg}). 

We propose two complementary implementation schemes to experimentally access the different phases, in particular the QAH phase, according to the phase diagrams stablished in Fig.~\ref{fig:fig2}: In realization (i) it suffices to a illuminate in a homogeneous, time-independent ("always-on") way the Lieb lattice with the Rydberg dressing laser, and to choose parameters such that inter-site distances $r_{AB} = r_{AC} = a$ and $r_{BC} = \sqrt{2} a$ are smaller than $r_c$. This realizes the regime where interactions $V_2$ are of comparable strength as $V_1$, $V_2 \lessapprox V_1$ (dashed line in Fig.~\ref{fig:fig2}b). Together with a non-zero staggering potential $\epsilon_A \neq 0$, this allows one to access QAH region (green parameter stripe), independently of its exact quantitative localization in the phase diagram. This is a striking difference of the realization of the topological phase in the Lieb lattice geometry, as compared to other schemes, including our proposal of Ref.~\cite{Dauphin_2012}, where the experimental accessibility according to the mean-field theory analysis requires very strong $V_2 \gtrapprox V_1$ next-to-nearest-neighbor interactions. Alternatively, thanks to the special geometry of the Lieb lattice, one can even access the extreme regime of finite (vanishing) interactions $V_2$ ($V_1$), corresponding to the vertical axis in the phase diagram of Fig.~\ref{fig:fig2}a in the absence of a staggering potential $\epsilon_A = 0$. This interaction pattern can be generated if the Rydberg dressing laser itself is applied according to a 45$^\circ$ tilted lattice structure in the xy-plane, so that atoms at B/C (A) lattice sites are exposed to a finite (vanishing) dressing light intensity, which results in finite (vanishing) interaction values $V_2$ ($V_1$). Finally, we remark that for the experimental detection of the topological phase, a variety of techniques have been proposed \cite{Alba2011,Price2012,Dauphin_2013,Wang2013,Hauke_2014,Deng_2014}, including schemes for the detection of edge states \cite{Goldman2012,Goldman_2013}, which can be adapted to the proposed setup.

\emph{Acknowledgments}:  A. D. thanks the F.R.S.-FNRS Belgium for financial support and N. Goldman and P. Gaspard for support and valuable discussions. We acknowledge support by the Spanish MINECO grant FIS2012-33152, the CAM research consortium QUITEMAD+ S2013/ICE-2801, and the U.S. Army
Research Office through grant W911NF-14-1-0103.

\appendix

\section{\label{app:mfh}Mean Field Hamiltonian}

In this Appendix, starting from Eq.~(\ref{eq:hamint}), we outline the derivation of Eq.~(\ref{eq:hamintmf}). Applying the Hartree-Fock decoupling to $n_in_j$, the mean-field Hamiltonian reads
\begin{equation}
H_\text{MF}=E_0+\sum_{\mathbf{k}}\Psi^\dagger_k \mathcal{H}_\mathbf{k} \Psi_\mathbf{k}\text{,}
\end{equation}
where 
\begin{equation}
\begin{split}
E_0/N_\lambda=&4V_1 \vert \xi_{AB}\vert^2+4V_2 \vert \xi_{BC} \vert^2\\
&-2V_1 \bar{n}_A(\bar{n}_B+\bar{n}_C)-4V_2 \bar{n}_B\bar{n}_C
\end{split}
\end{equation}
is a real number depending only on the expectation values of the operators, $\Psi_\mathbf{k}^\dagger=(c^\dagger_{A,\mathbf{k}},c^\dagger_{B,\mathbf{k}},c^\dagger_{C,\mathbf{k}})$ and where
\begin{equation}
\mathcal{H}_{\mathbf{k}}=\left(\begin{array}{ccc}
\mathcal{H}_\mathbf{k}^{11}& -2J' \cos{k_x} & -2J' \cos{k_y} \\
-2J'^* \cos{k_x} & \mathcal{H}_\mathbf{k}^{22}& -4 V_2|\xi_{BC}|f_{\mathbf{k},\theta}\\
-2J'^* \cos{k_y} & -4 V_2|\xi_{BC}|f^*_{\mathbf{k},\theta} & \mathcal{H}_\mathbf{k}^{33}\end{array}\right)\text{,}
\end{equation}
with $J'=J+\xi_{AB}V_1$, $f_{\mathbf{k},\theta}=\cos{k_x}\cos{k_y}\cos{\theta}-i\sin{k_x} \sin{k_y} \sin{\theta}$ and 
\begin{equation}
\begin{split}
&\mathcal{H}_\mathbf{k}^{11}=2(\bar{n}_B+\bar{n}_C)V_1\text{,} \\
&\mathcal{H}_\mathbf{k}^{22}= 2\bar{n}_AV_1+4V_2\bar{n}_C\text{,}\\
&\mathcal{H}_\mathbf{k}^{33}=2\bar{n}_AV_1+4V_2\bar{n}_B\text{.}
\end{split}
\end{equation}
We introduce the CDW order parameter $\rho_1=\bar{n}_A-(\bar{n}_B+\bar{n}_C)$, the nematic order parameter $\rho_2=\bar{n}_B-\bar{n}_C$ and separate the real $\xi_{BC}^\text{R}$ and imaginary $\xi_{BC}^\text{I}$ contributions to the next-to-nearest neighbors hopping amplitude. This allows us to write the real number $E_0$ as $E_0/N_\lambda=4V_1 \vert \xi_{AB}\vert^2+4V_2 ((\xi_{BC}^\text{R})^2+(\xi_{BC}^\text{I})^2)-V_1/2(n^2-\rho_1^2)-V_2(1/4(n-\rho_1)^2-\rho_2^2)$ and the matrix elements as 
\begin{equation}
\begin{split}
&\mathcal{H}_{\mathbf{k}}^{AA}=(n-\rho_1)V_1\text{, } \\
&\mathcal{H}_{\mathbf{k}}^{BB}=(n+\rho_1)V_1+(n-\rho_1-2\rho_2)V_2\text{, }  \\
&\mathcal{H}_{\mathbf{k}}^{CC}=(n+\rho_1)V_1+(n-\rho_1+2\rho_2)V_2\text{, } \\
&\mathcal{H}_{\mathbf{k}}^{AB}=-2J' \cos{k_x} \text{, } \mathcal{H}_{\mathbf{k}}^{AC}=-2J' \cos{k_y} \text{, }\\
&\mathcal{H}_{\mathbf{k}}^{BC}=-4V_2 \xi_{BC}^\text{R} \cos{k_x}\cos{k_y}+4 V_2\xi_{BC}^\text{I}\text{i} \sin{k_x} \sin{k_y} \text{.}
\end{split}
\end{equation}
We emphasize that the contribution proportional to $\rho_1$ in the next-to-nearest neighbor interaction is pathological and leads to an unstable free-energy: physically, the next-to-nearest neighbor interaction does not see A sites and therefore cannot generate an imbalance between sites $A$ and $B,C$. We therefore omit this term in the remainder.

\section{\label{app:nnre}Generation of real nearest-neighbor (NN) hopping renormalization and real next-to-nearest-neighbor (NNN) hopping from interactions}
In this section, we discuss the value of the renormalization of the NN hopping amplitude due to the NN interaction and the generation of the NNN hopping due to the NNN interaction. We show that these assumptions are reasonable by comparing it to the solution of the full MF equations on a chosen axis.  

\begin{figure}[t]
\begin{center}
	\includegraphics[scale=1]{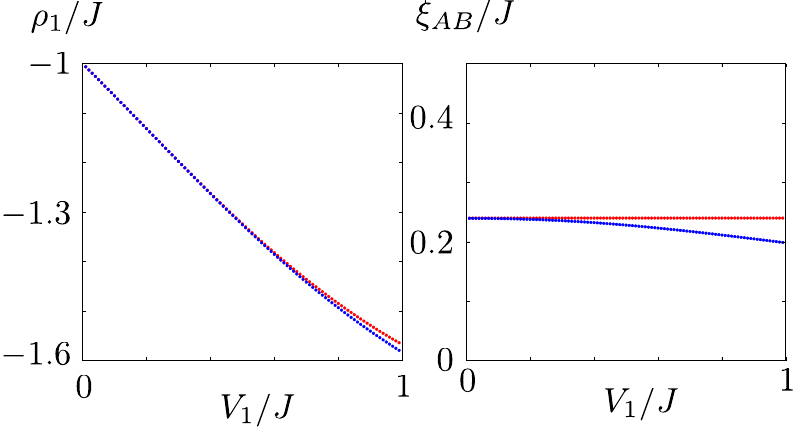}
	\caption{Order parameters $\rho_1$ and $\xi_{AB}$ when $V_2=0$ for the full mean field Hamiltonian in blue and for the mean field Hamiltonian with the approximation $\xi_{AB}=0.24J$ in red. The maximal error due to the approximation is of the order of $0.05J$.}
\label{fig:app1}
\end{center}
\end{figure}

Firstly, the NN interaction renormalizes the NN hopping amplitude $J \rightarrow J'=J+\xi_{AB}V_1$. This term will not lead to new phases in the phase diagram, but affects the position of the different phases. It is thus important to take it into account. To this end, we fix the value of $\xi_{AB}$ as the solution of a simplified mean-field equation \cite{Weeks_2010,Dauphin_2012}:  starting from the mean field Hamiltonian defined in the App.~\ref{app:mfh}, we fix $\rho_1=\rho_2=\xi_{BC}^\text{R}=\xi_{BC}^\text{I}=0$. The free energy is therefore given by the analytical expression:
\begin{equation}
\begin{split}
F/N_\Lambda=&4V_1 \xi_{AB}^2\\
&-\frac{1}{\pi^2}\int_\text{BZ} \sqrt{4(J+V_1 \xi_{AB})^2(\cos^2{k_x}+\cos^2{k_y})}\text{,}
\end{split}
\end{equation}
and the minimization of the free energy leads to the equation for $\xi_{AB}$:
\begin{equation}
\xi_{AB}=\frac{1}{\pi^2}\int_\text{B.Z.} \frac{J\sqrt{\cos^2 k_x+\cos^2 k_y}}{4}\simeq 0.24J\text{.}
\end{equation}
Fig.~\ref{fig:app1} compares the value of the order parameters along the axis $V_2=0$ for the full mean field Hamiltonian (blue dots) and the mean filed Hamiltonian with the assumption $\xi_{AB}=0.24J$: the approximation agrees very well with the complete solution with a maximal absolute error of the order of $0.05J$, being indeed a small correction as compared to $J$. 

Secondly, the NNN interaction generates a real NNN hopping contribution $\xi_{BC}^\text{R}$. We also take it into account and fix its value as the solution of a simplified mean-field equation, taking the Hamiltonian without interaction and only the contribution to the real NNN hopping due to the Hartree Fock decoupling of the NNN interaction. The real term is therefore expressed as $E_0=4V_2(\xi_{BC}^\text{R})^2$ and $\mathcal{H}_\mathbf{k}$ is written as
\begin{eqnarray}
\mathcal{H}_\mathbf{k} = - \left(\begin{array}{ccc}0 & 2J \cos k_x & 2J \cos k_y \\ 2J \cos k_x  & 0 & J_2^{\text{R}} \cos k_x \cos k_y \\ 2 J \cos k_y & J_2^{\text{R}} \cos k_x \cos k_y & 0\end{array}\right)\nonumber\\
\end{eqnarray}
with $J_2^{\text{R}} = 4 V_2 \xi_{BC}^\text{R}$.

\begin{figure}[t]
\begin{center}
	\includegraphics[scale=1]{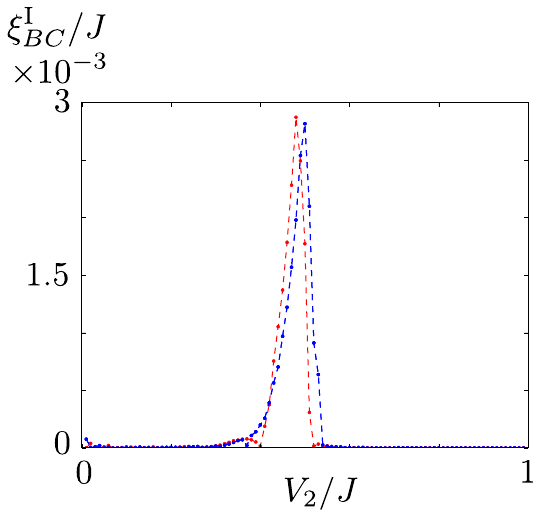}
	\caption{Order parameter $\xi_{BC}^\text{R}$ when $V_1=0$ for the full mean field Hamiltonian in blue and for the mean filed Hamiltonian with the approximation $\xi_{BC}^\text{R}=-0.1J$ in red.}
\label{fig:app2}
\end{center}
\end{figure}

In this case, the free energy needs to be minimized numerically. We find that $\xi_{BC}^\text{R}$ is almost constant and that its value is $\xi_{BC}^\text{R}=-0.1J$. Figure~\ref{fig:app2} compares the value of the order parameter $\xi_{BC}^\text{I}$ along the axis $V_1=0$ for the full mean field Hamiltonian (blue points) and for the mean field Hamiltonian under the assumption $\xi_{BC}^\text{R}=-0.1J$. The approximate curve fits well to the one of the full solution.

\section{\label{app:detailsmf}Details of the MF phase diagram}

\begin{figure}[t]
\begin{center}
	\includegraphics[scale=1]{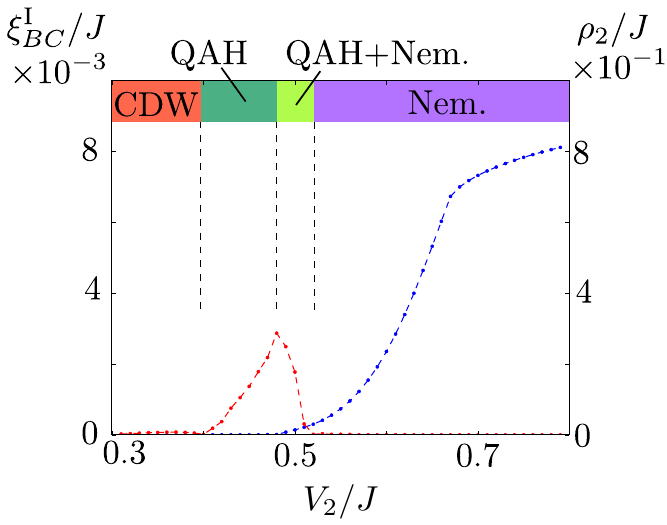}
	\caption{Order parameters $\xi_{BC}^\text{I}$ and $\rho_2$ on the vertical axis $V_1=0$. The order parameters are considered to be nonzero when they are greater than the numerical threshold $10^{-4}J$.}
\label{fig:app3}
\end{center}
\end{figure} 

In this Appendix, we discuss in detail the behavior along one line of the mean field phase diagram. Figure~\ref{fig:app3} shows the order parameters $\xi_{BC}^\text{I}$ and $\rho_2$ for $\epsilon_A=0$ and along the vertical axis $V_1=0$. The order parameters are considered to be $0$ if their values are smaller than the numerical threshold set to $10^{-4}J$. For small $V_2$, the system is in the CDW phase with $\rho_1\neq 0$. At $V_{2c,\text{QAH}}=0.37J$, $\xi_{BC}^\text{I}$ is becoming nonzero and the system enters the topological phase. To verify this last statement, we have computed the Chern number using the FHS algorithm on the effective Hamiltonian and found $C_1=0$, $C_2=1$, $C_3=-1$. At $V_2>V_{2c,\text{Nem.}}=0.47J$, $\rho_2$ becomes non zero. In this case, the nematic and the QAH phase coexist and interestingly the system is still in a topological regime with $C_2=1$. If $V_2$ is increased further, $\xi^\text{I}_{BC}$ continuously decreases to zero and at $V_2>V_{2c,\text{Nem. pure}}=0.52J$, the system enters in a purely ungapped nematic phase.

\section{\label{app:details_rydberg}Details on Rydberg dressing implementation}
\textit{Outline of the scheme and effective two-body interaction potential. --} The internal dynamics of the atoms in the Lieb lattice is described by the Hamiltonian $H = \sum_i H_i + \sum_{i<j} H_{ij}$, where the single-particle terms $H_i = (\Omega \ket{r}\bra{g}_i + h.c.) + \Delta \ket{r} \bra{r}_i$ describes the laser coupling (in dipole and rotating wave approximation) of atoms in the electronic ground state $\ket{g}$ to a Rydberg state $\ket{r}$, with $\Omega$ and $\Delta$ denoting the laser Rabi frequency and detuning, respectively. Pairwise interactions of atoms excited to $\ket{r}$ are given by $H_{ij} = V_{ij} \ket{r}\bra{r}_i \otimes  \ket{r}\bra{r}_j$. For Rydberg $s$-states with repulsive van-der-Waals interactions $V_{ij} > 0$ the isotropic interactions fall off quickly with the interparticle distance, $V_{ij} = C_6 |\mathbf{r}_i - \mathbf{r}_j|^{-6}$ \cite{gallagher-book}.

We are interested in far off-resonant laser coupling, $\Omega/\Delta \ll 1$, where atoms initially residing in the electronic ground state $\ket{g}$ 
obtain a small admixture of Rydberg character. The effective Born-Oppenheimer two-particle potential for a pair of dressed ground state atoms can be derived in fourth-order perturbation theory (see Ref.~\cite{Dauphin_2012} for details), yielding 
\begin{equation}
\label{eq:real_part}
(\Delta E)_{\ket{gg}} = 2 \frac{\Omega^4}{\Delta^3} \left[ 1+ \frac{2\Delta}{V_{ij}} \right]^{-1},
\end{equation}
which is shown in Fig.~\ref{fig:fig3}. Here, the trivial single-particle (fourth order in $\Omega/\Delta$) AC-Stark-shift has been subtracted. The Rydberg blockade radius corresponds to the length scale $r_c = (C_6 /2 \Delta)^{1/6}$, given by the condition $2 \Delta \approx V_{ij}$. It determines where the crossover from the plateau-like short-distance behavior, with a universal energy shift of $(\Delta E)_{\ket{gg}} = 2 \, \Omega^4/\Delta^3$ (up to a sub-leading correction), and the weakly interacting regime, with reduced van-der-Waals type interactions at large distances, $(\Delta E)_{\ket{gg}}  = (\Omega/\Delta)^4 V_{ij}$, takes place.

\textit{Experimental laser and atomic parameters and imperfections --}
Whereas an optimal set of parameters (Rydberg state, laser parameters) depend on specifics of the setup and the atomic species used, one can see that reaching realistic conditions is feasible: for a lattice spacing $ a \approx$ 500nm, Rydberg $s$ states of principal quantum number $n$ around $n=30$, with van-der-Waals interactions of the order of $2 \hbar \times$ 100 MHz, a detuning of $\Delta = 2 \pi \hbar \times$ 50 MHz and a Rabi frequency $\Delta = 2 \pi \hbar \times$ 5 MHz, the perturbative treatment in ($\Omega/\Delta = 0.1$) yields an effective interaction shift between ground state atoms of $(\Delta E)_{\ket{gg}} \approx 2\pi \times 10 \,\text{kHz}$. This is on the order of typical tunnelling rates of a few kHz. At the same time, it is much larger than effective decoherence rate $\gamma_{\rm eff} = (\Omega/\Delta)^2 \gamma_r = 2 \pi \times 50$\,Hz, which is induced by the finite lifetime of Rydberg states of a few tens of microseconds of $n=30$ $s$-states at room temperature \cite{gallagher-book,saffman-rmp-82-2313}. One can show that for the considered lattice fillings of order one (1/3 and 2/3) at the present parameters the effective interaction is well-described by pair-wise interactions, and higher-order many-body effects as well as mechanical effects between the laser-dressed ground state atoms are negligible (see also the discussion in Ref.~\cite{Dauphin_2012}).

\end{document}